\documentclass[twocolumn,showpacs,preprintnumbers,amsmath,amssymb,prl]{revtex4}

\usepackage{graphicx}
\usepackage{color}
\usepackage{dcolumn}
\usepackage{bm}
\usepackage{amsmath}
\def\degC{\kern-.2em\r{}\kern-.3em C}

\begin{document}
\preprint{******}

\title{Storage and Retrieval of a Squeezed Vacuum}

\author{Kazuhito Honda$^{1}$, Daisuke Akamatsu$^{2}$, Manabu Arikawa$^{1}$, Yoshihiko Yokoi$^{2}$, Keiichirou Akiba$^{2}$, \\
Satoshi Nagatsuka$^{2}$, Takahito Tanimura$^{2}$, Akira Furusawa$^{3}$, and Mikio Kozuma$^{1,2,4}$}

\affiliation{%
$^{1}$Interactive Research Center of Science, Tokyo Institute of Technology, 2-12-1 O-okayama, Meguro-ku, Tokyo 152-8550, Japan}

\affiliation{%
$^{2}$Department of Physics, Tokyo Institute of Technology, 2-12-1 O-okayama, Meguro-ku, Tokyo 152-8550, Japan}

\affiliation{%
$^{3}$Department of Applied Physics, School of Engineering, The University of Tokyo, 7-3-1 Hongo, Bunkyo-ku, Tokyo 113-8656, Japan}

\affiliation{%
$^{4}$PRESTO, CREST, Japan Science and Technology Agency, 1-9-9 Yaesu, Chuo-ku, Tokyo 103-0028, Japan}%

\date{\today}
\begin{abstract}
Storage and retrieval of a squeezed vacuum was successfully 
demonstrated using electromagnetically induced transparency.
930~ns of the squeezed vacuum pulse 
was incident on the laser cooled $^{87}$Rb atoms 
with an intense control light in a coherent state. 
When the squeezed vacuum pulse was slowed 
and spatially compressed in the cold atoms, 
the control light was switched off. 
After 3~$\mu$s of storage, 
the control light was switched on again and the squeezed vacuum was retrieved, 
as was confirmed using the time-domain homodyne method.
\end{abstract}

\maketitle

Storing photonic information in the atomic ensemble is one of the most important tasks to implement complicated protocols related to quantum repeaters or local networking.  Experimental research into quantum memory has primarily followed two paths. The first is a method based on electromagnetically induced transparency (EIT)\cite{Harris1997,Phillips-2001,Liu-2001,Akamatsu-2004,Kuzmich-2005,Eisaman-2005}, and the second involves the use of quantum non-demolition measurement (QND) of spins\cite{Polzik-2004,Schori-2002}. Storage of the single photon state was successfully demonstrated with the EIT using nonclassical lights generated from atoms\cite{Kuzmich-2005,Eisaman-2005} or those created by the parametric process\cite{AKiba-2007}. Storage for continuous variable of light was also demonstrated with spin QND using coherent state of light, where the obtained fidelity between the incident and the retrieved lights was significantly higher than the limit for classical recording\cite{Polzik-2004}. However, no quantum memory has been reported up to now for non-classical continuous variable of light. In the present paper, we report on the successful demonstration of storage and retrieval of the squeezed vacuum using cold ensemble of Rb atoms as a memory medium, where quadrature noise of the retrieved light was monitored with the homodyne method and sub-shot noise feature was clearly observed.

\begin{figure}
    \includegraphics[width=0.95 \linewidth]{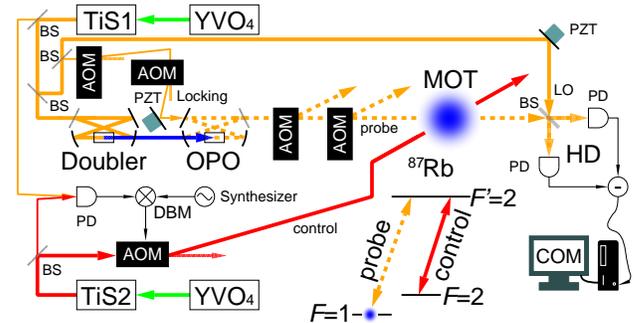}
  \caption{\label{F_setup} (Color online)
Schematic diagram of the experimental apparatus.
The inset shows the energy diagram for $^{87}$Rb used for the EIT.
YVO$_4$: frequency-doubled YVO$_4$ laser;
TiS: Ti:sapphire laser;
BS: beam splitter;
PD: photo detector;
LO: local oscillator;
HD: homodyne detector; 
DBM: double balanced mixer;
AOM: acoustic-optical modulator;
COM: computer for data acquisition and analysis.
}%
\end{figure}

%
Our experimental setup is shown schematically in Fig.\ref{F_setup}. We used laser cooled $^{87}$Rb atoms as an EIT medium, since high transmittance can be easily achieved in the EIT\cite{Arikawa-2007} and all atoms can be hyperfine pumped before starting the experiment. In the experiment using a hot glass cell, fresh atoms always enter the beam area and emit photons through the hyperfine pumping process, which may lead to unwanted elevation of shot noise level.

All of the data in our experiment were obtained through the cycle which consisted of the preparation 
of a cold atomic sample in $5^2$S$_{1/2}\:F=1$ state (9 ms) and the measurement period (1 ms) (about the detail, see \cite{Arikawa-2007}). 
During the measurement period, 
the EIT was observed with the squeezed vacuum, where the optical depth of the cold atomic sample was approximately five.
The probe and control transitions were 
$F=1 \leftrightarrow F'=2$ and $F=2 \leftrightarrow F'=2$ in $D_1$ line.
The beat signal between two Ti:sapphire lasers (TiS1 and TiS2) was mixed with the synthesizer output and the beat frequency was reduced enough to drive an acoustic-optical modulator (AOM). The frequency difference between the light from TiS1 and the control light generated by diffracting the output of TiS2 with the AOM was stabilized to hyperfine splitting of two ground states by this feed-forwarding method\cite{feed-forward}.
%

%
As the probe light, 
the squeezed vacuum was generated with a sub-threshold optical parametric oscillator (OPO) driven by the second harmonic light from a doubler, where periodically-poled KTiOPO$_4$ crystals were utilized\cite{pp-KTP}. The quadrature amplitude of the generated squeezed vacuum was detected using a balanced optical homodyne detector (HD). 

Power drift of a local oscillator (LO) directly affects the precision of the shot noise level. Since three hours were taken to acquire the data, the LO power was monitored and actively stabilized, where the standard deviation for the LO power drift during three hours was suppressed to 0.004 dB (0.1\%).

Usually a LO beam has high frequency classical power noises and there is also slight power imbalance between two LO beams incident on the two photodetectors of a homodyne system, which means experimentally obtained shot noise is slightly larger than the real one. To minimize such a difference, we first applied classical power modulation to the LO and compensated the modulation signal from the HD by changing the reflectivity of the half beam splitter, which was carried out by adjusting the angle of the splitter. Eventually we could achieve $-58$~dB of suppression for the classical modulation. Next, we measured the classical power noise of the LO beam. We cut off one of the LO beams without classical modulation in front of the photodetector and measured the variation of the noise level. The classical noise level of the LO beam averaged over the bandwidth from 1~MHz to 2~MHz was approximately $+3$~dB against the shot noise. These results mean the difference between experimentally obtained shot noise level and the real one was extremely small. 

In order to analyze the quadrature noise of the probe field, 
we employed the time-domain 
method\cite{homodyne,neergaard-nielsen:083604,takei:060101,Arikawa-2007}, 
where we acquired the signals 
from the homodyne detector using a computer 
and performed Fourier transform of the obtained real-time waveforms. 
We hereinafter represent the measured power noise as $S(\theta)$, 
where $\theta$ is the phase difference between the squeezed vacuum and the LO. 
Note that we define the relative phase $\theta$ such that 
$S(\theta)$ is maximized (minimized) at $\theta=\pi/2$ ($0$).
%

%
In the experiment, the relative phase $\theta$ 
between the squeezed vacuum and the LO was actively stabilized 
using a weak coherent state of light from TiS1 (locking beam) and two PZTs during the cold atom preparation period. The technical detail is described in \cite{Arikawa-2007}.
Just before the depumping stage, 
the locking light was shut off using two AOMs in front of the OPO, 
and the feedback voltages to the PZTs were held, 
which enabled us to observe the squeezed vacuum 
that passed through the atoms in the absence 
of the locking light during the 1-ms measurement period. 
The same circular polarizations were utilized for both the squeezed vacuum and the control lights. The optical path of the squeezed vacuum was crossed with that of the control light with an angle of $2.5^\circ$ at the center of the atomic ensemble. 
The $1/e^2$-radii of the control light and the squeezed vacuum 
at the ensemble were 550 $\mu$m and 150 $\mu$m, respectively.

\begin{figure}
  \begin{center}
    \includegraphics[width=.38\textwidth]{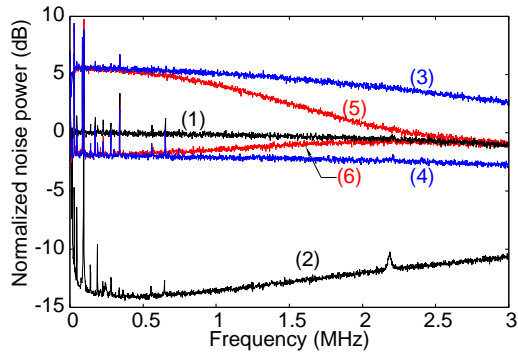}
  \end{center}
  \caption{\label{F_EIT} (Color online)
Frequency spectra of the quadrature noise for the squeezed vacuum.
(1): shot noise;
(2): electric noise from the homodyne detector;
(3): quadrature noise at $\theta=\pi/2$, $S(\pi/2)$;
(4): quadrature noise at $\theta=0$, $S(0)$;
(5),(6): $S(\pi/2)$ and $S(0)$ of the squeezed vacuum 
transmitted through the EIT medium.
}%
\end{figure}

%
The noise spectra of the squeezed vacuum are shown in Fig. \ref{F_EIT}, 
where the real-time homodyne signal was acquired 
with a sampling rate of $5\times 10^7$ samples$/$s.
Vertical resolution of the digital oscilloscope was 8 bits 
and thus the dynamic range for the power measurement was almost 50 dB. 
The obtained Fourier-transformed power noise was averaged over 1,000 trials. 
Trace (1) indicates the shot noise, 
and Trace (2) indicates the thermal noise of the detector, 
where the sharp peaks in the frequency region below 700 kHz 
were due to the electric noise from the environment. 
Traces (3) and (4) indicate $S(\pi/2)$ and $S(0)$ 
of the squeezed vacuum without the cold atoms. 
The levels of $S(\pi/2)$ and $S(0)$ with respect to the shot noise 
were approximately 6 dB and $-2$ dB, respectively. Squeezing level was limited mainly due to the gray track of the nonlinear crystal. 
Traces (5) and (6) indicate $S(\pi/2)$ and $S(0)$ of the squeezed vacuum, 
which passed through the EIT medium. 
The amplification and attenuation of the quadrature noises 
were observed within some limited bandwidth, 
which is a clear feature of the EIT \cite{Arikawa-2007}. 
The full bandwidth at half maximum of both $S(\pi/2)$ and $S(0)$ 
were estimated to be 2.7 MHz for the control light power 
of 0.3 mW.

\begin{figure}
  \begin{center}
    \includegraphics[width=.4\textwidth]{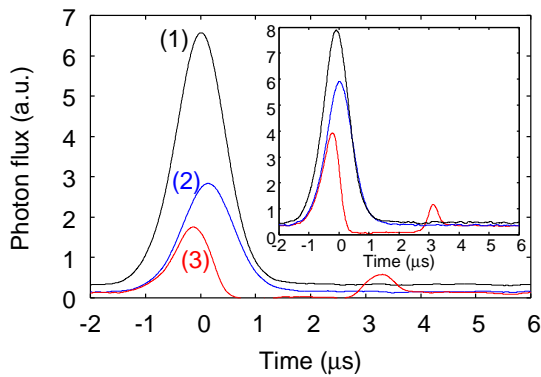}
  \end{center}
  \caption{\label{F_Flux} (Color online)
Temporal variations of the photon fluxes of the squeezed vacuum pulses analyzed by the method I.
The inset shows those obtained by the method II.
In the case of the storage,
the control light was shut off at 0~$\mu$s and switched on at 3~$\mu$s.
(1): original pulse;
(2): delayed pulse;
(3): stored and retrieved pulse.
}%
\end{figure}

%
%
Next, we performed the storage and retrieval of the pulsed squeezed vacuum by dynamically changing the control light intensity. 
In our experiment, 930 ns of the squeezed vacuum pulses were created from the continuous squeezed vacuum using two AOMs.
In order to prevent unwanted optical loss, 
we used a 0th-order (non-diffracted) beam as the probe pulse, 
rather than a 1st-order diffracted light\cite{Akamatsu-2007}. 
As a result of the finite diffraction efficiency of the AOMs, 
there was a residual photon flux at the tail of the pulse, 
which was 5\% of the peak of the pulse. 
We injected the squeezed vacuum pulse into the cold atoms. When the squeezed vacuum pulse was compressed in the atoms due to the slow propagation, we shut off the control light inducing the transparency 
and the state of the light was mapped onto the atoms. 
Switching the control light on after 3~$\mu$s of storage, 
the light having the original pulse information was retrieved. 
It is noted that the relative phase between the retrieved light and the LO is determined by that between the control light and the LO in a frame rotating with the frequency of hyperfine splitting. The employment of the feed-forwarding method is thus inevitable.
One sequence consisted of 11~$\mu$s, 
and we repeated the same sequences 90 times 
during the 1-ms measurement period. 
Measurement was repeated $10^4$ times, 
and the power noise for the quadrature amplitude 
was averaged over all of the sequences. 
The sampling rate was $2\times10^8$~samples$/$s.

In order to evaluate the temporal variation of the quadrature noise, we employed two different methods. In the first method (hereinafter called method I), Fourier-transformation was performed for each 640 ns of the time window, which corresponded to the time resolution of this analysis. Here, the quadrature noise with the bandwidth of 1 MHz and the center frequency of 1.5 MHz was evaluated. In the second method (hereinafter called method II), a temporal function was estimated from the storage experiment with a coherent state of light, which was $f(t-t_0)=\exp[-(t-t_0)/\tau]$ and $\tau=250$ ns. The obtained homodyne signal was multiplied by $f(t-t_0)$ and was integrated from $t_0$ to $t_0+750$ ns so that the quadrature signal having the temporal mode of $f(t-t_0)$ was evaluated \cite{PolzikSinglePhoton}. Advantage of the method I is that contribution of electric noise from the environment can be neglected because only the high frequency component is used for the analysis. Disadvantage is that the squeezing is degraded, since high frequency component experienced optical loss due to the limited EIT window (see Traces (5) and (6) in Fig.2). In contrast, the second method has an advantage that the squeezing level extracted from the real time homodyne signal can be maximized, since the temporal function determining the quadrature mode is matched by that of the retrieved pulse. It is noted that Fourier transform of $f(t-t_0)$ is Lorentzian with the bandwidth of 640kHz. Therefore, the squeezing in the low frequency region is included in this analysis. Since we have the whole spectrum of our homodyne system (Fig.2), in the second analysis we excluded specific frequency components where classical spike noises appeared.

%

%
We first confirmed that the photon flux of the squeezed vacuum was slowed, 
stored, 
and retrieved. 
In general, 
an averaged photon number is proportional to $S(\theta)+S(\theta+\pi/2)-2$, 
when $S(\theta)$ is normalized with the shot noise. 
Figure \ref{F_Flux} shows temporal variation 
of the photon flux analyzed by the method I. 
Trace (1) indicates the photon flux 
of the original squeezed vacuum pulse in the absence of the cold atoms. 
Trace (2) corresponds to the squeezed vacuum pulse 
that passed through the cold atoms under the EIT, where the pulse was delayed for 135 ns compared to the original squeezed vacuum pulse\cite{Akamatsu-2007}. 
Trace (3) indicates the photon flux 
of the squeezed vacuum pulse when the control 
light was shut off at 0 $\mu$s and re-injected after 3 $\mu$s.  
While the height of the retrieved photon flux 
was only 20\% of that of the delayed pulse, 
it was still higher than the residual photon fluxes 
at the tails of the delayed and original pulses. 
After retrieval, 
the tail of the signal gradually matched the photon flux of the delayed pulse. 
This is reasonable because the atomic state was considered 
to approach the steady state of the EIT 
constructed by the residual squeezed vacuum 
and the control light after the retrieval of the stored pulse.
The inset of Fig. 3 shows the photon fluxes obtained by the method II. 
Since the squeezing in the low frequency region is analyzed, photon flux under the EIT is increased.  

\begin{figure}
  \begin{center}
    \includegraphics[width=.38\textwidth]{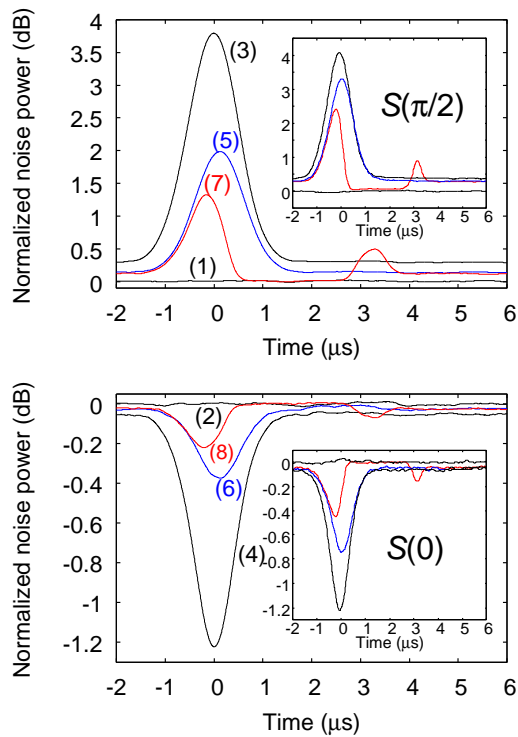}
  \end{center}
  \caption{\label{F_Store} (Color online)
Temporal variation of the quadrature noises, 
$S(\pi/2)$ and $S(0)$ obtained by the method I.
The insets are those analyzed by the method II.
The upper diagram is for $S(\pi/2)$ and
the lower diagram is for $S(0)$.
The power noise levels were normalized with the shot noise.
(1) and (2): shot noise;
(3) and (4): original pulses;
(5) and (6): delayed pulses;
(7) and (8): stored and retrieved pulses.
}%
\end{figure}

%
The main purpose of our experiment is to show 
that phase sensitive quadrature noise can be stored and retrieved using the EIT. 
Figure \ref{F_Store} shows the temporal variations 
of quadrature noises $S(\pi/2)$ and $S(0)$ obtained by the method I, 
where Traces (1) and (2) are the shot noise. 
Traces (3) and (4) are the $S(\pi/2)$ and $S(0)$ 
of the original pulse in the absence of the cold atoms. 
Traces (5) and (6) show $S(\pi/2)$ and $S(0)$ of the pulse 
delayed by the EIT and Traces (7) and (8) 
are those when storage and retrieval were performed. 
While the control light was shut off, 
the noise levels matched the shot noise 
because the squeezed vacuum was completely absorbed 
by the optically dense atomic medium. 
When the control light was switched on again, quadrature noise levels suddenly changed and $0.489 \pm0.007$ dB (+11.9\%) of the anti-squeezing and $-0.070 \pm 0.006$ dB ($-1.6$\%) of the squeezing were obtained, where both the statistical error and the standard deviation of the LO power drift were included. The insets of Fig.4 represent the results obtained by the method II, where $0.90 \pm 0.01$ dB (+23\%) of the anti-squeezing and $-0.16 \pm 0.01$ dB ($-3.6$\%) of the squeezing were extracted.
While these values are remarkably small, 
they clearly exceed the values of the tails of the delayed pulse, 
as well as that of the original pulse. 
Optimizing the temporal shape of the incident squeezed vacuum will help to increase the retrieval efficiency and improve the squeezing level\cite{LukinOptimize}.
The anti-squeezing and squeezing for the incident light estimated by the method II were $4.10 \pm 0.01$ dB (+157\%) and $-1.24 \pm 0.01$ dB ($-24.8$\%), respectively. While fidelity between the incident and the retrieved lights can be calculated from these results \cite{TakeiSqueez}, such a discussion is misleading. The fidelity approaches 1 as the squeezing of the incident light decreases, because the state was not displaced from the vacuum \cite{Namiki}.

We independently measured the squeezing of the retrieved lights for 2 $\mu$s and 4 $\mu$s of the storage durations. Increasing the storage time by 2 $\mu$s, the retrieved squeezing (the offset was subtracted) decreased by half. We also measured the residual magnetic field caused by the eddy current. Atomic coherence time determined by inhomogeneous broadening of the ground states was estimated to be the order of 10 $\mu$s. We do not have a clear explanation about this discrepancy.

%
In conclusion, 
we have succeeded in storing and retrieving the squeezed vacuum with the EIT. 
Note that when the squeezed vacuum is stored in the atomic ensemble, the collective atomic spins are also squeezed and their quantum fluctuation is suppressed 
below the standard quantum limit\cite{Akamatsu-2004,Lukin-2002,Ueda-1993}. 
Performing a single photon count for the field 
retrieved partially from the squeezed atoms, 
a highly nonclassical atomic state will be 
generated\cite{Hau-1999-cat,neergaard-nielsen:083604,Wakui-2006}.

%

%
%
The authors would like to thank M. Ueda, N. Takei, 
and K. Usami for several stimulating discussions. 
Two of the authors (D. A. and K. A.) were supported by the Japan Society for the Promotion of Science. 
This study was supported by a Grant-in-Aid for Scientific Research (B), 
the Promotion of Science for Young Scientists, 
and the 21th Century COE Program at Tokyo Tech, 
`Nanometer-Scale Quantum Physics', by MEXT.

After submission of this manuscript, we learned that related
work is being pursued \cite{Lvovsky2007}.

\bibliography{manuscript}

\end{document}